\documentclass[useAMS,usenatbib,epsfig]{mn2e}
\newif\ifAMStwofonts
\AMStwofontstrue
\usepackage{amsmath,amsfonts,epsfig,natbib}

\def\reff@jnl#1{{\rm#1\/}}

\def\aj{\reff@jnl{AJ}}                  
\def\araa{\reff@jnl{ARA\&A}}            
\def\apj{\reff@jnl{ApJ}}                        
\def\apjl{\reff@jnl{ApJ}}               
\def\apjs{\reff@jnl{ApJS}}              
\def\ao{\reff@jnl{Appl.Optics}}         
\def\apss{\reff@jnl{Ap\&SS}}            
\def\aap{\reff@jnl{A\&A}}               
\def\aapr{\reff@jnl{A\&A~Rev.}}         
\def\aaps{\reff@jnl{A\&AS}}             
\def\azh{\reff@jnl{AZh}}                        
\def\baas{\reff@jnl{BAAS}}              
\def\gca{\reff@jnl{GeCoA}}              
\def\jrasc{\reff@jnl{JRASC}}            
\def\memras{\reff@jnl{MmRAS}}           
\def\mnras{\reff@jnl{MNRAS}}            
\def\pra{\reff@jnl{Phys.Rev.A}}         
\def\prb{\reff@jnl{Phys.Rev.B}}         
\def\prc{\reff@jnl{Phys.Rev.C}}         
\def\prd{\reff@jnl{Phys.Rev.D}}         
\def\prl{\reff@jnl{Phys.Rev.Lett}}      
\def\pasp{\reff@jnl{PASP}}              
\def\pasj{\reff@jnl{PASJ}}              
\def\qjras{\reff@jnl{QJRAS}}            
\def\skytel{\reff@jnl{S\&T}}            
\def\solphys{\reff@jnl{Solar~Phys.}}    
\def\sovast{\reff@jnl{Soviet~Ast.}}     
\def\ssr{\reff@jnl{Space~Sci.Rev.}}     
\def\zap{\reff@jnl{ZAp}}                        
\def\nat{\reff@jnl{Nature}}             

\title[VSA observations of the SZ effect in the Corona Borealis Supercluster]{A VSA search for the 
extended Sunyaev--Zel'dovich Effect in the Corona Borealis Supercluster}

\author[R. Genova-Santos et al.] {Ricardo Genova-Santos,$^1\thanks{E-mail:
  rgs@iac.es}$ Jos\'e Alberto Rubi\~no-Martin,$^1$ Rafael Rebolo,$^{1,4}$ Kieran 
\newauthor Cleary,$^2\thanks{Present address: Jet Propulsion Laboratory, 4800 Oak Grove Drive, 
Pasadena, CA 91109, USA}$
Rod D. Davies,$^2$ Richard J. Davis,$^2$
Clive Dickinson,$^2\thanks{Present address: California Institute of Technology, 1200 E. 
California Blvd., Pasadena, CA 91125, USA}$ Nelson Falc\'on,$^{1,5}$ 
\newauthor Keith Grainge,$^3$ Carlos M. Guti\'errez,$^1$ Michael P. Hobson,$^3$ 
Michael E. Jones,$^3\thanks{Present address: Astrophysics Group, Dept. of Physics, Keble Road, Oxford, OX1 3RH, UK}$
\newauthor R\"udiger Kneissl,$^3\thanks{Present address: Dept. of Physics, University of California at Berkeley, 
CA 94720-7300, USA}$ 
Katy Lancaster,$^3\thanks{Present address: Astrophysics Group, University of Bristol, 
Tyndall Avenue, Bristol BS8 1TL, UK}$ Carmen P. Padilla- Torres,$^1$ Richard D.E. 
\newauthor Saunders,$^3$ Paul F. Scott,$^3$ Angela C. Taylor$^3\S$ and 
 Robert A. Watson$^2\thanks{Present address: 
  Instituto de Astrof\'{i}sica de Canarias, 38200 La Laguna, Tenerife, Spain}$  \\
$^1$ Instituto de Astrofis\'{i}ca de Canarias, 38200 La Laguna, Tenerife, Canary Islands, Spain \\
$^2$ Jodrell Bank Observatory, University of Manchester, Macclesfield, Cheshire, SK11 9DL, UK \\
$^3$ Astrophysics Group, Cavendish Laboratory, University of Cambridge CB3 OHE, UK \\ 
$^4$ Consejo Superior de Investigaciones Cient\'{\i}ficas, Spain \\ 
$^5$ Dpt. de F\'{\i}sica, FACYT, Universidad de Carabobo, Venezuela \\} 
\date{Accepted Received In original form}
\pagerange{\pageref{firstpage}--\pageref{lastpage}}
\pubyear{2005}

\begin{document}

\label{firstpage}
\maketitle

\begin{abstract}
We present interferometric imaging at $33$~GHz of the Corona Borealis supercluster, using the 
extended configuration of the Very Small Array. A total area of $24$~deg$^2$ has been imaged, with an 
angular resolution of $11$~arcmin and a sensitivity of $12$~mJy/beam. 
The aim of these observations is to search for Sunyaev--Zel'dovich (SZ) detections from known clusters of 
galaxies in this supercluster 
and for a possible extended SZ decrement due to diffuse warm/hot gas in the intercluster 
medium. Hydrodynamical simulations suggest that a significant part of the missing baryons in the local 
Universe may be located in superclusters. 
 
The maps constructed from these observations have a significant contribution from primordial 
fluctuations. We measure negative flux values in the positions of the ten richest clusters in the region. 
Collectively, this implies a 3.0-sigma detection of the SZ effect. For two of these clusters, A2061 and A2065, 
we find decrements of approximately 2-sigma each. 

Our main result is the detection of two strong and resolved 
negative features at $-70\pm12$ mJy/beam ($-157\pm27~\mu$K) and $-103\pm10$~mJy/beam 
($-230\pm23~\mu$K), 
respectively, located in a region with no known clusters, near the centre of the supercluster. 
We discuss their possible origins in terms of primordial CMB anisotropies 
and/or SZ signals related to either unknown clusters or to a diffuse extended warm/hot gas distribution. Our 
analyses have revealed that a primordial CMB fluctuation is a plausible explanation for the weaker feature 
(probability of 37.82\%). For the stronger one, neither primordial CMB (probability of 0.33\%) nor SZ can 
account alone for its size and total intensity. The most 
reasonable explanation, then, is a combination of both primordial CMB and SZ signal. Finally, we explore what 
characteristics would be required for a filamentary structure consisting of warm/hot diffuse gas in order to 
produce a significant contribution to such a spot taking into account the constraints set by  X-ray data.

\end{abstract}

\begin{keywords}
cosmology:observations -  cosmic microwave background - galaxies:clusters - 
techniques:interferometric
\end{keywords}

\section{Introduction}

The mean baryon density of the Universe 
is one of the most relevant cosmological parameters, as it influences baryonic 
structures on all scales, from the abundances of primordial nuclei to the large-scale 
distribution of galaxies and intergalactic gas. Indeed, its value is also a prediction 
of the standard Big Bang model that may be tested observationally. 
The calculation of the primaeval abundances of light elements, such as deuterium,
combined with standard Big Bang nucleosynthesis, allows a very precise
determination of this parameter for standard models \citep{burles_01a}:
$\Omega_B=(0.020 \pm 0.002)~h^{-2}=0.038 \pm 0.004$, where here 
$h\equiv h_{100} = H_0/100~km~s^{-1}Mpc^{-1} = 0.73$ is adopted
for the last term. Likewise \citet{rauch_97a} derive a lower limit 
$\Omega_B = 0.021~h^{-2}=0.042$ for a $\Lambda$CDM model from observations of the
Ly$\alpha$ forest absorption in a selected sample of seven high resolution quasar
spectra at $z=2$. More recently, the best-fit cosmological model to 
the first year \textit{WMAP} data release also gives a value 
$\Omega_B=(0.0224 \pm 0.0009)~h^{-2}=0.044 \pm 0.002$ \citep{spergel_03}, while 
the recent VSA results give 
$\Omega_B=(0.0234^{+0.0012}_{-0.0014})~h^{-2}=0.0464^{+0.0024}_{-0.0028}$ 
\citep{rebolo_04}.

In principle, the consistency between these three completely 
independent methods of estimating the baryon density is straightforward, but at 
$z=0$ in the present day Universe, the sum over all the well observed components 
give a considerably smaller value. Indeed, \citet{fukugita_98a} made an estimate 
of the global budget of baryons in all known states, namely the different 
kinds of stars, the atomic and molecular gas, and the plasma in galaxy clusters 
and in groups. They infer a value 
$\Omega_B=(0.010 \pm 0.003)~h^{-2}=0.020 \pm 0.007$.

Thus, if no serious errors have been made in the theoretical estimates, which seems 
unlikely given the agreement between the different methods, it seems that most of the 
baryons are yet to be detected in the present 
day Universe. This is the well-known ``baryon problem''.

One of the most important hypotheses related to this hidden matter is
based upon hydrodynamical simulations \citep{cen_99a,dave_01a} that predict 
the formation at low redshift
($z<1$) of a very diffuse gas phase with temperatures $10^{5} < T < 10^{7}$~K, which
is neither low enough to have permitted condensations into stars or to form cool 
galactic gas, nor as high as that of the hot gas present in galaxy clusters. Such
low density gas could account for a substantial fraction of the 
missing  baryons, and according to these simulations it should be distributed in
large scale 
sheet-like structures and filaments connecting clusters of galaxies, due to the 
infall of baryonic matter into previously formed dark matter filaments. Moreover, 
this amount of gas would not violate the constraints on the spectral distortions of 
the cosmic microwave background (CMB). 
It constitutes what is known as the ``warm/hot intergalactic 
medium'' (WHIM) and would be observed in the soft X-ray band. The detection of its 
radiation could be obscured by the presence of many galactic foregrounds and by 
extragalactic contributions from groups of galaxies, clusters or AGNs. 
Nevertheless, several attempts in recent years have been made to detect it, and 
some detections have been claimed, either by studying the correlation 
between the observed soft X-ray structures and selected galaxy overdense 
regions \citep{scharf_00a,zappacosta_02a} or by detecting a soft X-ray excess in
clusters of galaxies \citep{finoguenov_03a}, or in their proximity 
\citep{briel_95a,tittley_01a,soltan_02a}.

As indicated by \citet{cen_99a}, both X-ray emission and the thermal Sunyaev--Zel'dovich 
(SZ) effect can be cross-correlated with galaxy or galaxy
cluster catalogues to search for the missing baryons. The SZ effect typically originates 
from the richest clusters of galaxies that contain extended 
atmospheres of hot gas ($k_B T_e \sim 10~keV$). But there may be other objects 
that could also produce a detectable SZ signal, such as superclusters of galaxies, where 
even though low baryon overdensities are expected, path lengths may be long so that 
a significant SZ effect could build up since the effect is proportional to the line 
of sight integral of the electron density \citep{birkinshaw_99}. Also, it is 
reasonable to look at 
superclusters as optimal regions for the detection of WHIM, since
simulations show that this gas should be distributed in filamentary structures 
extending over several tens of Mpc and connecting clusters of galaxies. In fact, 
\citet{molnar_98} looked in the \textit{COBE}-DMR data in the region of the Shapley 
supercluster, but found no clear evidence of an extended SZ effect and therefore  
set an upper limit of $\la -100~\mu$K on the gas temperature. On the other hand, several 
studies have been carried out not by searching in selected overdense regions, but by 
trying to obtain a statistical SZ detection from intra-supercluster (ISC) gas 
from analyses of the whole sky, first using the \textit{COBE}-DMR data \citep{banday_96a} and, 
more recently, the first release of \textit{WMAP} data 
\citep{fosalba_03a,monteagudo_04,myers_04a,monteagudo_04b}. This is 
mainly carried out by correlating CMB maps with 
templates built from galaxy or galaxy cluster catalogues. 

In this study we concentrate on the Corona Borealis supercluster (CrB-SC). This 
supercluster does not contain high flux density radio sources which may inhibit SZ 
measurements. The supercluster members have relatively high X-ray fluxes 
\citep{ebeling_98a,ebeling_00a} which suggests that detectable SZ effects could build 
up. Also, the angular size of the supercluster core is suitable for mosaiced 
observations with the VSA extended array, which has a primary beam FWHM of $2.1\degr$. 
Taking these factors into account, we selected this supercluster as the most interesting 
object of its type within the region of sky observable by the VSA 
(which have a declination coverage of $\approx -7\degr$ to $+63\degr$).

In Section~2 we present a general description of the CrB-SC
and summarize previous observational work. Section~3 includes an overview of the VSA 
interferometer and explains the data reduction and the map making procedure.
In Section~4 we discuss the SZ effects from the individual clusters in the region 
and the possible causes of the two observed negative features. Conclusions are 
presented in Section~5.

\section{The Corona Borealis Supercluster}

\begin{figure}
\includegraphics[width=\columnwidth]{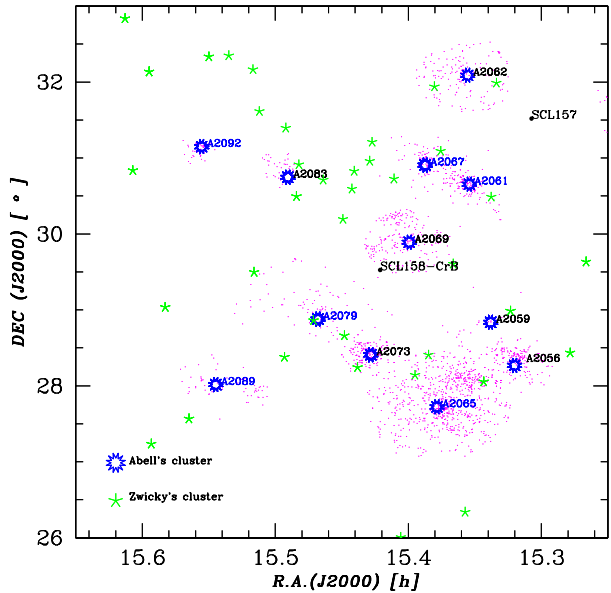}
\caption{Spatial distribution of the Abell and Zwicky clusters in the CrB-SC core region. The Abell
clusters labelled with blue letters belong to the CrB-SC. Also indicated are the positions of the 
SCL168 (CrB) and SCL157 supercluster centres, as reported by \citet{einasto_01a}. Red dots indicate 
the positions of the galaxies at redshifts within $\pm0.02$ of the supercluster redshift and 
inside the optical radius of each Abell cluster. (The redshifts and the optical radius are taken 
from the NASA Extragalactic Database.)}
\label{fig:cor_plot}
\end{figure}

The CrB-SC
is one of the most prominent examples of superclustering
in the northern sky. \citet{shane_54a}, after counting galaxies on the Lick 
Observatory photographic plates, were the first to remark upon the extraordinary 
cloud of galaxies that constitute the supercluster, and later \citet{abell_58a} 
noted the presence of this concentration of clusters of galaxies including it in 
his catalogue of second order clusters. Depending on author, the number of
clusters belonging to this SC ranges from six to eight 
\citep{postman_88a,small_97a}, but we will focus on the classification given in 
the \citet{einasto_01a} catalogue, according to which the CrB-SC 
includes eight clusters, around the position $R.A.=15^{h}25^{m}16.2^{s}$,
$DEC=+29\degr 31'30''$, at a redshift $z\approx~0.07$. These clusters are listed 
in Table~\ref{tab:clusters}, along with their characteristics.  
Six of these 
(A2061, A2065, A2067, A2079, A2089 and A2092) are located in the core of the SC, in 
a $\approx 3\degr \times~3\degr$ region, while there are two others (A2019 and A2124)
at an angular distance of $\approx 2.5\degr$ from the core. \citet{einasto_01a} 
include only A2061 and A2065 as X-ray emitting clusters. There are four other Abell 
clusters (A2056, A2005, A2022 and A2122) in this region at redshifts around 
$z~\approx~0.07$, two (A2069 and A2083) at $z~\approx~0.11$ and
another two (A2059 and A2073) that are even more distant. There are also several Zwicky
clusters but without redshift measurements. The spatial distribution of these 
clusters can be seen in Figure~\ref{fig:cor_plot}.

The first dynamical study of the CrB-SC was carried out by 
\citet{postman_88a}, through the study of a sample of 1555 galaxies in the
vicinity of Abell clusters. They report 97 new redshift measurements over 
the previous 85. They conclude that the masses of all 
clusters in the core of the CrB-SC lie in the range 
$1.5-8.9~\times~10^{14}~M_{\odot}$, while the mass of the SC is
approximately $\approx 8.2~\times~10^{15}~M_{\odot}$, which is probably enough to bind
the system. They proved also that the dynamical timescales are comparable with the
Hubble time, making it unlikely that the system could be virialised, as one might  
expect. This work was extended by \citet{small_98a} by increasing the
number of galaxies with known redshifts to 528.
They quote a value for the mass of the supercluster of 
$3-8~\times~10^{16}\ h^{-1}~M_{\odot}$, a value slightly higher than the previous
one. This is because \citet{postman_88a} made the assumption that the
mass-to-light ratio in the CrB-SC is the same as in the richest Abell clusters 
and they used a supercluster volume three times smaller. They remark that almost
one third of the galaxies in the region are not linked to any Abell cluster, and
also emphasize the great contribution to the projected surface density of galaxies
of the background cluster A2069 and its surrounding galaxies, located at a redshift 
$z~\approx 0.11$, suggesting the existence of the so-called ``A2069 supercluster''.
On the other hand, \citet{marini_04} have analysed on the \textit{BeppoSax} X-ray data 
in a search for evidence of a merging signature between the pairs of clusters candidates 
in the region of CrB-SC A2061-A2067 and A2122-A2124. They find no 
clear evidence of interaction but detect a candidate shock inside A2061. 
\begin{table*}
\caption{Clusters: basic data. The coordinates are the optical centre of each cluster. The last 
two columns show the predicted flux which would be observed 
by the VSA for those clusters with X-ray measurements (labelled ``pred'', see Section~4.1 for details), 
and the measured value in our final maps (labelled ``VSA''). The error bars contain the thermal 
noise ($\sigma_n$), the average primordial CMB (which has been computed to be 
$\sigma_{CMB}\approx 19$~mJy/beam) and the residual sources ($\sigma_{sour}\approx 6$~mJy/beam) 
added in quadrature.}
\label{tab:clusters}
\begin{center}
\begin{tabular}{@{}cccccccc}
\hline
Cluster& RA$^{(a)}$ & DEC$^{(a)}$ & z$^{(a)}$ & $L_{x}$(0.1-2.4~keV) & $T_{e}$& $\Delta
S_{\nu}^{pred}$&$\Delta S_{\nu}^{VSA}$ \\
        & (J2000) &  (J2000)  &     & ($10^{44}~h_{50}^{-2}~ergs^{-1}$) & (keV)  & (mJy/beam) & (mJy/beam) \\
\hline
A2019   & 15 02 57.2  & +27 11 17 & 0.0807 &              &             &       &$ -2.6\pm24.5$\\
A2061   & 15 21 15.3  & +30 39 17 & 0.0784 & $3.95^{(b)}$ & $5.6^{(b)}$ &$-54.7$&$-53.8\pm27.4$\\
A2065   & 15 22 42.6  & +27 43 21 & 0.0726 & $4.94^{(b)}$ & $8.4^{(b)}$ &$-61.7$&$-42.8\pm23.4$\\
A2067   & 15 23 14.8  & +30 54 23 & 0.0739 & $0.86^{(c)}$ & $3.1^{(c)}$ &$-27.7$&$-26.5\pm27.4$\\
SCL-158 & 15 25 16.2  & +29 31 30 &        &              &             &    	&	      \\
A2079   & 15 28 04.7  & +28 52 40 & 0.0690 &              &             &       &$ -6.7\pm22.6$\\
A2089   & 15 32 41.3  & +28 00 56 & 0.0731 &              &             &       &$-28.9\pm27.9$\\
A2092   & 15 33 19.4  & +31 08 58 & 0.0669 &              &             &       &$-21.0\pm28.5$\\
A2124   & 15 44 59.3  & +36 03 40 & 0.0656 & $1.35^{(b)}$ & $3.7^{(b)}$ &$-44.3$&$-19.4\pm29.2$\\
\hline
A2069   & 15 23 57.9  & +29 53 26 & 0.1160 & $8.92^{(b)}$ & $7.9^{(b)}$ &$-57.1$&$-15.2\pm27.4$\\
A2073   & 15 25 41.5  & +28 24 32 & 0.1717 & $3.74^{(c)}$ & $5.6^{(c)}$ &$-29.9$&$-33.4\pm22.6$\\
\hline
\end{tabular}
\medskip

Notes: (a) NASA Extragalactic Database,  (b) \citet{ebeling_98a},  (c) \citet{ebeling_00a}.\\ 
Note the use of $h_{50} = H_0/50~km~s^{-1}Mpc^{-1}$.
\end{center}
\end{table*}

\section{Observations and data reduction}

\subsection{Observations with the VSA interferometer}

The Very Small Array (VSA) is a 14-element heterodyne interferometric array, 
tunable between 26 and 36~GHz with a 1.5~GHz
bandwidth and a system temperature of approximately 35~K. It is located at an 
altitude of 2400~m at the Teide Observatory in Tenerife. 
For this study, the observing frequency was set at 33~GHz, and we
used the extended configuration, which uses 
conical corrugated horn antennae with 322~mm 
apertures, and has a primary beam FWHM of $2.1 \degr$, and a
synthesised beam FWHM $\approx 11$~arcmin at the observing frequency. For a 
detailed description of the instrument see e.g. \citet{VSApaperI}.

Situated next to the main array is a two-element interferometer, 
which consists of two 3.7-m diameter dishes with a north--south
baseline of 9~m, giving a 
resolution of $4$~arcmin in a $9$~arcmin field. This interferometer is 
used for source subtraction (SS), monitoring radio sources simultaneously with 
the main array observations, as described in \citet{VSApaperII}. This 
strategy neatly copes with the additional problem of source variability.

The observations were carried out during the period spring 2003 -- summer 2004,
although most of the data were taken in the first months, while the last months 
were used
only to improve the noise levels in those fields with a poorer signal to noise, 
and to obtain a better measurement of the fluxes of some of the radio sources.
Initially, we observed five individual patches covering the interesting central 
regions of the supercluster, plus two additional pointings separate from the 
main mosaic and centred on clusters A2019 and A2124. We later added four more 
pointings to improve the coverage of the supercluster core. 
These
observations are described in Table~\ref{tab:observations}. Each daily observation  
was approximately four hours in duration. We dedicated between 5 (for CrB-D and 
CrB-G) and 37 (for CrB-H) days to each pointing. We quote both total observation 
and integration times, the latter indicating the data retained after flagging 
(for instance, periods of bad weather).

\begin{table*}
\begin{minipage}{150mm}
\begin{center}
\caption{Summary of the observations with the VSA. We list the
central coordinate of the eleven pointings, the total observation time, the
integration time and the thermal noise achieved for each map 
(computed from the map far away from the primary beam).}
\begin{tabular}{lccccc}
\hline
Pointing & RA (J2000) & DEC (J2000)&$\rm T_{obs}$&$\rm T_{int}$ & Thermal noise \\
   &            &            & (hr)  & (hr) & (mJy/beam)      \\ 
\hline
CrB-A & 15 23 12.00 & +28 06 00.0 & 54 & 50 &  12.4 \\      
CrB-B & 15 27 48.00 & +29 24 00.0 & 70 & 70 &  10.8 \\      
CrB-C & 15 22 48.00 & +30 21 00.0 & 33 & 33 &  18.9 \\       
CrB-D & 15 32 00.00 & +30 45 00.0 & 19 & 19 &  20.5 \\      
CrB-E & 15 32 00.00 & +28 18 00.0 & 22 & 22 &  19.7 \\       
CrB-F & 15 02 57.20 & +27 11 17.3 & 47 & 43 &  14.4 \\      
CrB-G & 15 45 00.00 & +36 03 57.6 & 19 & 19 &  21.4 \\      
CrB-H & 15 23 00.00 & +29 13 30.0 &167 &130 &  10.2 \\
CrB-I & 15 27 24.00 & +30 33 00.0 & 56 & 41 &  18.6 \\      
CrB-J & 15 32 00.00 & +29 31 30.0 & 55 & 39 &  18.9 \\      
CrB-K & 15 28 00.00 & +28 12 00.0 & 41 & 33 &  20.3 \\      
\hline
\end{tabular}
\label{tab:observations}
\end{center}
\end{minipage}
\end{table*}

\subsection{Calibration and data reduction} 

The primary calibrator for the VSA is Jupiter. 
The flux scale is transferred to other calibration sources: Tau A (the Crab Nebula) and 
Cas A, which are observed daily. A detailed description of the VSA calibration process is 
presented in \citet{VSApaperI}, whereas in \citet{VSApaperV} and \citet{dickinson_04} are 
remarked the specifications adopted for the VSA extended configuration.

In the first VSA studies (e.g. \citet{VSApaperI,VSApaperV}) the calibration scale was 
based on the effective temperature of Jupiter given by \citet{mason_99}, which 
extrapolated to $33$~GHz is $T_{Jup}=153\pm5~K$ ($3$\% accuracy in temperature). In the 
most recent VSA published data \citep{dickinson_04} we rescale our results using the 
Jupiter temperature as derived from the \textit{WMAP} first-year data: $T_{Jup}=146.6\pm2.0~K$
\citep{page_03}, which agrees with the \citet{mason_99} value at $1\sigma$ level, but 
reduces the calibration error from $3$ to $1.5$\% in temperature terms ($6$\% to $3$\% in 
the power spectrum). In this paper we have also rescaled the Jupiter temperature to the 
\textit{WMAP} value, which is the most accurate value published to date.

The correlated signal from each of 91 VSA baselines is processed
as described in \citet{VSApaperI} and \citet{VSApaperV}, and the  
data checks so described are also applied. 
 
\begin{table*}
\begin{minipage}{150mm}
\begin{center}
 \caption{List of all the radio sources with measured fluxes at $33$~GHz above $50$~mJy. 
 We quote the extrapolated flux at $33~$GHz, and the estimated flux at $33~$GHz using 
 the SS observations, which have been used to subtract the sources from the data.} 
 \label{tab:sources}
  \begin{tabular}{|c|c|c|c|c|c|c|}
   \hline
& Name  & RA (J2000) & DEC (J2000)   & Extrapolated flux & Measured flux by	\\
&       & 	   &		 &  NVSS-GB6 at 33~GHz (mJy) &  the SS at 33~GHz (mJy) \\
     \hline
CrB-F   & 1459+2708 & 14 59 39.60 & 27  8 16.0 &  83  &	 $132\pm12$   \\
        & 1504+2854 & 15  4 27.30 & 28 54 25.0 &  76  &	 $193\pm14$   \\
        & 1509+2642 & 15  9 39.10 & 26 42 45.0 &  39  &	 $ 58\pm10$   \\
\hline
\noalign{\smallskip}
Mosaic  & 1514+2931 & 15 14 20.90 & 29 31  9.0 & 104  &  $ 95\pm 6$   \\ 
        & 1514+2855 & 15 14 40.30 & 28 55 39.0 &  72  &  $ 63\pm 8$   \\ 
        & 1514+2943 & 15 14  3.70 & 29 43 21.0 &  33  &  $ 50\pm13$   \\ 
        & 1521+3115 & 15 21  1.80 & 31 15 50.0 & 196  &  $ 69\pm43$   \\ 
        & 1522+3144 & 15 22  9.50 & 31 44 18.0 & 243  &  $296\pm 7$   \\ 
        & 1522+2808 & 15 22 48.90 & 28  8 51.0 & 133  &  $ 97\pm 5$   \\ 
        & 1527+3115 & 15 27 18.20 & 31 15 14.0 & 294  &  $157\pm 6$   \\ 
        & 1528+3157 & 15 28 52.60 & 31 57 34.0 &  75  &  $ 98\pm 7$   \\ 
        & 1529+3225 & 15 29 38.70 & 32 25 23.0 &  60  &  $ 84\pm38$   \\
        & 1531+2819 & 15 31 21.40 & 28 19 26.0 &  30  &  $ 58\pm 9$   \\  
        & 1532+2919 & 15 32 20.20 & 29 19 40.0 &  71  &  $ 51\pm12$   \\ 
        & 1535+3126 & 15 35 58.90 & 31 26 25.0 &  47  &  $ 53\pm 5$   \\ 
        & 1537+2648 & 15 37  6.40 & 26 48 24.0 &  23  &  $ 56\pm18$   \\ 
        & 1539+3103 & 15 39 15.90 & 31  3 59.0 & 107  &  $ 86\pm15$   \\ 
        & 1539+2744 & 15 39 38.80 & 27 44 33.0 & 244  &  $218\pm10$   \\ 
\hline
\noalign{\smallskip}
CrB-G   & 1538+3557 & 15 38 57.40 & 35 57  9.0 &  37  &  $ 54\pm12$   \\   
        & 1540+3538 & 15 40 31.70 & 35 38 26.0 &  24  &  $ 52\pm16$   \\ 
        & 1544+3713 & 15 44 44.70 & 37 13 22.0 &  25  &  $ 50\pm11$   \\ 
        & 1546+3631 & 15 46  7.60 & 36 31  7.0 &  26  &  $ 51\pm18$   \\ 
        & 1546+3644 & 15 46 38.30 & 36 44 30.0 &  44  &  $ 59\pm15$   \\ 
        & 1547+3518 & 15 47 51.70 & 35 18 59.0 &  87  &  $ 77\pm12$   \\ 
        & 1552+3716 & 15 52  5.30 & 37 16  5.0 &  47  &  $ 50\pm18$   \\ 
\hline   
\end{tabular}
\end{center}
\end{minipage}
\end{table*}

\subsection{Source subtraction}

In order to remove the effect of radio sources from the
data, we followed a slightly different approach to
that considered for the primordial CMB fields (see \citet{cleary_04}).
For those fields, the Ryle Telescope (RT) was used to identify all the sources
above a given threshold at 15~GHz (see \citet{VSApaperII}
and \citet{VSApaperV} for the compact and extended arrays, respectively).
Using this catalogue, the source subtraction baseline monitored all
these sources in real time, and the
derived fluxes were subtracted from the VSA main array data.
However, for the present study a complete scanning of the region with the
RT was not available owing to observing time constraints, so in order to 
build a source catalogue in the region, we proceeded as follows. 
All the sources in the NVSS--1.4~GHz \citep{condon_98a} and
GB6--4.85~GHz \citep{gregory_96a} catalogues closer than $2\degr$ from each
pointing position were identified.
Using the fluxes from these catalogues, the inferred spectral index
between the respective frequencies was used to extrapolate the flux to 
$33$~GHz. Finally, all sources with predicted fluxes above $20$~mJy
at $33$~GHz were monitored using the SS baseline.
The flux threshold was chosen according to the requirements for the primordial 
CMB fields, but, as we shall see below, for the purposes of this
paper this limit can be relaxed.

Source positions were assigned according to the coordinates
in the GB6 catalogue. This catalogue has a lower angular resolution
(FWHM $\approx 3.5$~arcmin) than the NVSS (FWHM $\approx 45$~arcsec),
so multiple sources in NVSS could be associated with a single entry in the
GB6 catalogue. This was taken into account when computing
the spectral index for each source: the adopted flux at the NVSS frequency
was derived using the GB6 resolution, combining all sources in NVSS which
correspond to a given entry in the GB6 catalogue (i.e. for each GB6 source, 
we found all the NVSS sources closer than $5$~arcmin, applied the primary beam correction, 
and summed the corrected fluxes).

Taking into account that the
sensitivity of the GB6 catalogue is $18$~mJy at $4.85$~GHz,
together with a maximum value of the rising spectral index between
$1.4$~GHz and $33$~GHz (e.g. \citet{mason_99}) of $\alpha=0.5$
($S\propto\nu^{\alpha}$),
we are confident that we have measured all radio sources with fluxes above
$\approx 50$~mJy at $33$~GHz. Nevertheless, we appreciate that sources with inverted spectra and higher 
fluxes may exist, and identify four compact features at the $3\sigma_n$ level 
(being $\sigma_n$ the standard deviation of the thermal noise) in our maps of the CrB-SC region. 
We suspected that these may be radio sources not picked up by our strategy, but source 
subtractor observations suggest otherwise. No sources with detectable fluxes 
(i.e. $S > 20 $~mJy at the 2$\sigma$ level) were found at those positions.

A total of $74$ sources closer than $2\degr$ to any pointing were 
found to have an extrapolated flux greater than
$20~$mJy at 33~GHz, and of these, only $25$ 
had a measured flux with the source subtractor greater than $50~$mJy. 
In Table~\ref{tab:sources} we present the final list of identified sources with 
measured fluxes greater than $50$~mJy at $33$~GHz. These values were used to 
carry out the source subtraction.

It should be noted that this flux limit is larger than that used for the analyses 
of the primordial CMB fields with the extended VSA array. For that study, we used a flux limit of  
$20~$mJy, which guarantees that the residual contribution from unsubtracted sources is 
less than the flux sensitivity \citep{VSApaperV,dickinson_04,cleary_04}, and this could be 
achieved thanks to the RT survey at $15~$GHz.
However, as we shall see below (Section~$4.2$), the main result of this paper is
the detection of two anomalous cold regions with the VSA in the CrB-SC
region, that seem difficult to explain as primordial CMB features. 
We took our list of extrapolated fluxes and used the values to determine the effects on our 
measurements of the cold spots. We found that the individual effect of any one source was 
sufficient only to affect our measurements by less than $2$\%, and that the collective 
effect of all sources would cause effects of $1.3$\% and $3.2$\% in the two spots 
respectively. This latter error is only twice the calibration error. 
Thus for the purposes of this paper (in contrast with the VSA power spectrum measurements, 
where more stringent source constraints were required 
\citep{scott_03,VSApaperV,dickinson_04}), our source subtraction limit of $50~$mJy 
is sufficient.

In any case, an estimate of the confusion noise due to unresolved sources
below our source subtraction threshold can be made using the best fit 
power-law model obtained from the source count above $\approx 50$~mJy (see
fluxes in Table~\ref{tab:sources}),

\begin{equation}
n(S)=(17.6\pm2.3)\left(\frac{S}{70~mJy}\right)^{-2.28\pm0.83}mJy^{-1}sr^{-1}~.
\label{sourcecounts}
\end{equation}

Note that this fit has a slope compatible with that of the model  
presented in \citet{cleary_04}, which was obtained using the source count
from the VSA primordial CMB observations. However, the amplitude in our
case is a factor $\approx1.7$ larger.  This is as expected since there may
be a higher density of radio sources in the supercluster due to the
presence of sources associated with the member clusters. Indeed, the model
of \citet{cleary_04} predicts just $15$ radio sources above $50$~mJy in
our region of observation, whereas we find $25$. Using the model
described by equation~\ref{sourcecounts} we estimate that the
residual sources may introduce a confusion noise  
$\sigma_{sour}=7.3$~mJy/beam. Note however that this result may
overestimate the real confusion noise in the regions far away from
the CrB-SC clusters, where the source count should be described by the 
model given in \citet{cleary_04}.

\begin{figure*}
\includegraphics[width=9cm]{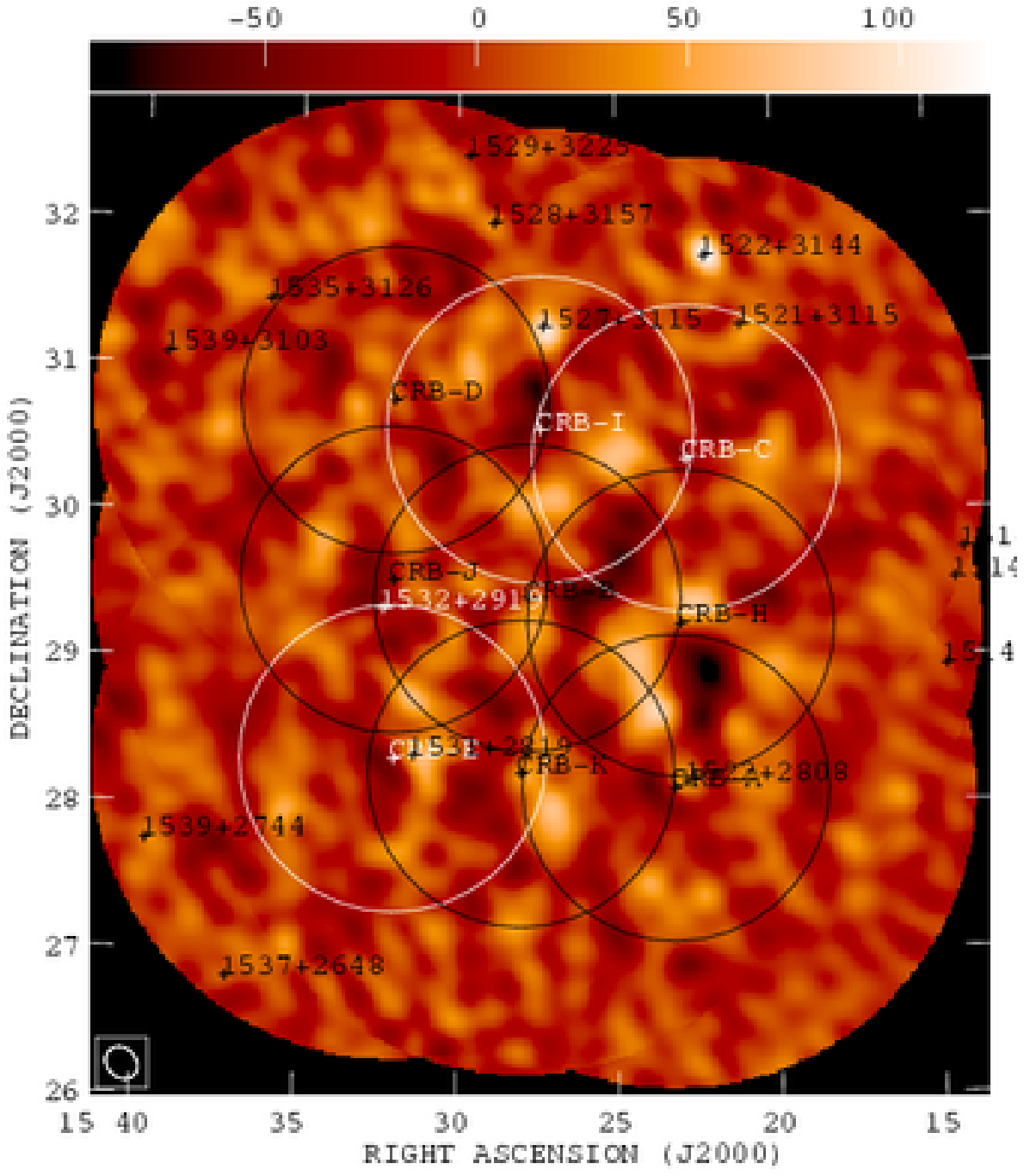}%
\includegraphics[width=9cm]{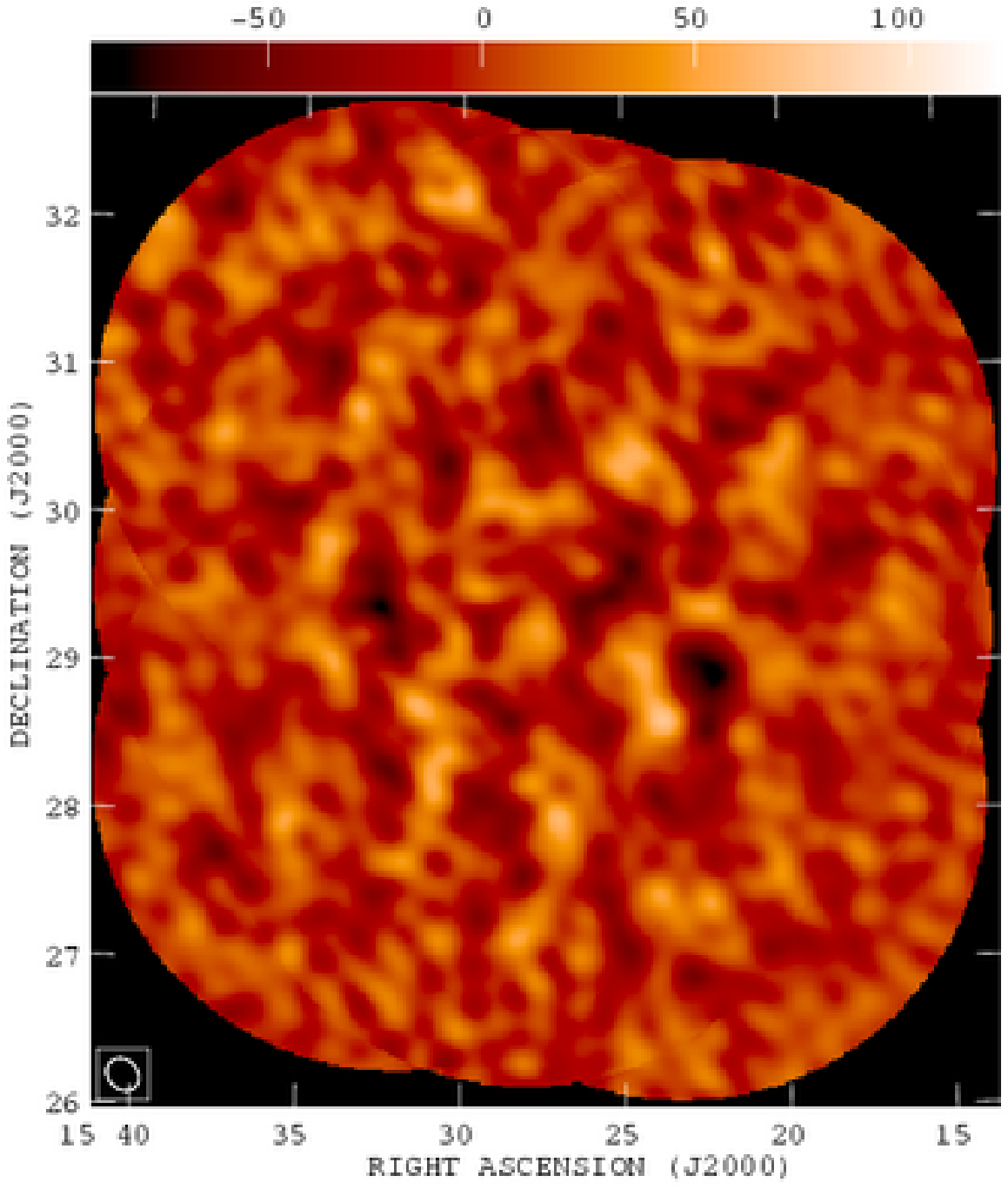}
\caption{\textsc{clean}ed VSA mosaics built up from pointings CrB-A,B,C,D,E,H,I,J,K. In the
un-source-subtracted mosaic (left panel) the circumferences indicate the primary beams 
FWHM ($2.1\degr$) of the nine 
pointings, and the crosses the positions of the monitored sources with measured fluxes above 
$50~$mJy. The right panel shows the \textsc{clean}ed source-subtracted mosaic. The synthesised 
beam FWHM is shown in the bottom left corner ($\approx 11$~arcmin). The noise level is 
practically uniform across the mosaic at a level $\approx 12$~mJy/beam.}
\label{fig:mosaics}
\end{figure*}

To conclude this section, we also report a 
problem which was detected during a late stage in the data processing, 
related to SS measurements. Since the source subtractor continually slews 
between pointed observations 
of point sources inside the observed VSA fields, the phase stability
of the system has to be checked several times during an observing run,
as described in \citet{VSApaperI}.
To this end, we observe several interleaved strong calibration
sources close to the considered region on the sky in order to
track the phase stability of the system properly. 
We found that one of the calibration sources used in the
CrB-SC central mosaic was not as bright as predicted by the 
extrapolation and hence was not bright enough to produce a sufficiently accurate 
determination 
of the phase. Using Monte Carlo simulations with the known calibrator 
flux and the sensitivity of the measurement, it is easy
to show that a weak calibration source produces an average underestimate 
of the source flux. From these simulations, we could try to correct all the 
flux measurements, but this would require us to make an assumption about the
variability of each source.
Instead, we decided to use a new estimator of the flux of the source
that does not require phase calibration.

Let us introduce the estimator $E[S^2] = V_R^2 + V_I^2 - 2 \sigma^2$ where
$S$ is the flux of the source we are measuring, $V_R$ and $V_I$ the real and the 
imaginary parts of the visibility, and $\sigma$ the statistical noise of each 
measurement. By construction, this
estimator is unbiased ($\langle E[S^2] \rangle = \langle S^2 \rangle$) and is 
not sensitive to phase
error of the instrument. If the noise is Gaussian, then $E$ is distributed
as a (displaced) $\chi^2$ with two degrees of freedom.
However, we must note that for our purposes, this estimator can only be used for
the case of non-variable sources, because it provides $\langle S^2 \rangle$ during 
the whole
period of observation, and not $\langle S \rangle$, which is the quantity we are 
interested in. 
Thus, we finally decided to use the estimator $E'=\sqrt{E}$ to derive the
fluxes. This estimator is obviously defined only for positive values of $E$ and 
in principle is biased to high values of the flux if the signal-to-noise ratio 
of the measurement is small. This bias was computed and we find that, for 
the flux sensitivity of the instrument, it is significant only for values of the flux 
smaller than $50$~mJy. However, we have reduced this bias by averaging the 
square estimator, $E$, between adjacent measurements of the same source to get a 
positive value prior to application of the square root.
Summarizing, even if we do not have an accurate determination of the phase
of the instrument at the time of the observation for some of the sources,  
their fluxes can be recovered from our measurements using this new 
estimator, although the error bar on the final measurement will be larger
than the one obtained with the standard procedure. 

In addition, and as a consistency check, we re-observed 
the strongest sources in the mosaic, with estimated fluxes above $\approx 60$~mJy, 
but now using an appropriate calibrator.   
A variability study was then applied to all these sources following the same
criteria as in
\citet{cleary_04}, and variable sources were identified.
Only five sources were found to be variable.
For non-variable sources, the new flux was directly compared to 
our previous determination to check that our flux values are robust.
Final values for the fluxes are presented in Table~\ref{tab:sources}.

\begin{figure}
\includegraphics[width=\columnwidth]{./figures/coronab_ss_cln_clus_wmapcal_zoom_decrj.ps}
\caption{The central region of the \textsc{clean}ed source-subtracted mosaic shown 
in right panel of Figure~\ref{fig:mosaics}. We indicate the locations of the four 
negative decrements. We also show the
positions of the clusters extracted from NED. Five-pointed stars are used for the 
CrB-SC clusters, triangles for the other Abell clusters, squares for the Zwicky clusters, 
and diamonds for the clusters present in other catalogues. The centre of the supercluster is 
indicated with a pentagon.
Solid contours show positive flux values, while dashed contours show negative flux values, where 
all contours correspond to $1.5\sigma_n$.}
\label{fig:mosaic_zoom}
\end{figure}

\begin{figure*}
\includegraphics[width=7.5cm]{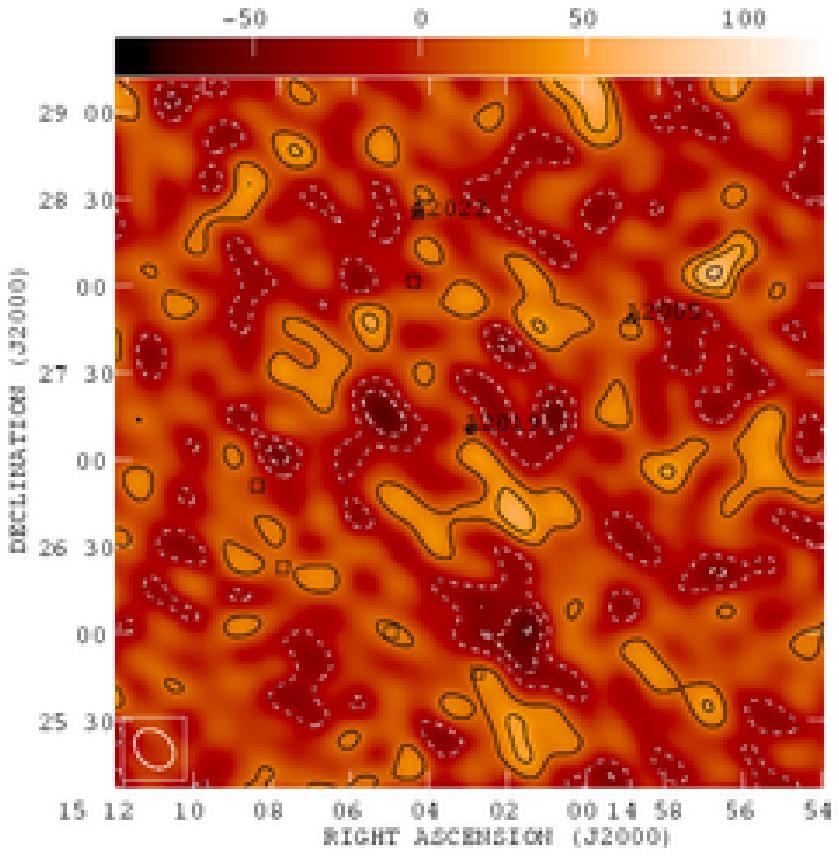}%
\includegraphics[width=7.5cm]{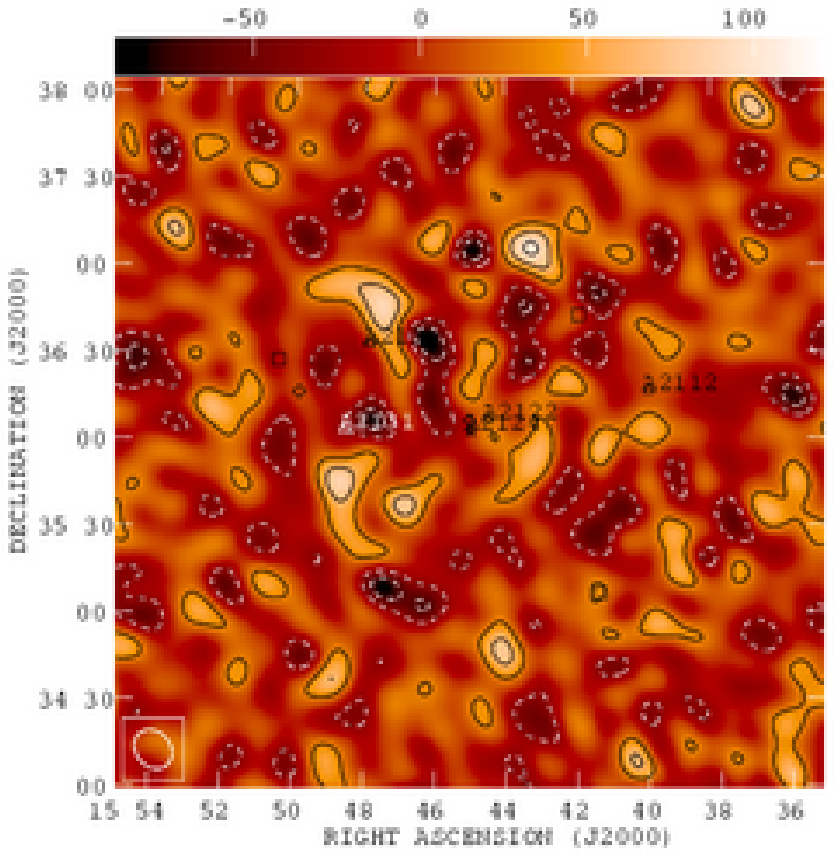}
\caption{\textsc{clean}ed and source-subtracted VSA maps corresponding to pointings 
CrB-F (left panel) and CrB-G (right panel). The nomenclature is the same as explained
for Figure~\ref{fig:mosaic_zoom}.}
\label{fig:maps}
\end{figure*}

\subsection{VSA maps}

Daily observations are calibrated and reduced as described in Section~3.2 
(and references therein) and are held as visibility files, which contain the real
and imaginary parts measured at each $uv$-position together with the associated rms
noise level. These files are loaded individually into \textsc{aips} 
\citep{greisen_94a}, where the map making process is carried out using standard 
tasks. The individual visibility files are stacked. Each of these stacks contains 
typically ${\simeq}~300\,000$ visibility points, each one averaged over 64 seconds. 
The source subtraction is implemented in the aperture plane using \textsc{aips} task 
\textsc{uvsub}, and then the maps for each 
individual pointing are produced. We used natural weighting, 
the most appropriate in this case since it produces the highest 
sensitivity and given that the sampling of the aperture plane is practically uniform.  
From the nine central pointings we produced the mosaiced map using standard 
\textsc{aips} tasks. \textsc{ltess} and \textsc{stess} were used to make a map in 
primary beam-corrected units, followed by a sensitivity map, which is practically 
uniform around the centre and fails in the outer regions. The two mosaics were 
divided to give the signal-to-noise, and the result multiplied by the central value 
of the sensitivity map in order to recover units of mJy/beam.
In order to avoid loss of signal to noise in the overlapping regions, and 
taking into account that the synthesised beams of the nine inner pointings have  
similar shapes and orientations, the \textsc{clean} mechanism 
(i.e. the deconvolution of the synthesised beam) has been applied directly to the mosaic, 
instead of to the individual pointings. To this end, we used the 
synthesised beam corresponding to the central pointing CrB-B. We placed 
\textsc{clean}-boxes around the strongest features, and \textsc{clean}ed down to a 
depth of $2\sigma_n$. 
In the case of the un-source-subtracted mosaics, 
\textsc{clean}-boxes were also placed around the identified radio 
sources. The same method was applied to the \textsc{clean} map 
corresponding to pointing CrB-F. In the case of the CrB-G map, as the signal-to-noise 
ratio is lower and it is not easy to disentangle the real 
features from the artefacts, we placed a \textsc{clean}-box around the 
region encompassing the primary beam FWHM, and \textsc{clean} was applied 
down to a depth of $3\sigma_n$ in this case.

The \textsc{clean}ed mosaics, before and after source subtraction, are shown in 
Figure~\ref{fig:mosaics}. In Figure~\ref{fig:mosaic_zoom} we present a larger scale plot 
of the regions of interest of the \textsc{clean}ed and source-subtracted mosaic.  
The maps resulting from the individual pointings 
CrB-F and CrB-G are shown in Figure~\ref{fig:maps}. The sensitivity values 
obtained in each individual pointing are shown in Table~\ref{tab:observations}, 
and the noise level of the overall mosaic is $12$~mJy/beam.

\section{Results and discussion}

\subsection{SZ effect from known clusters in the CrB region}

In Table~\ref{tab:clusters} we show the basic data of the eight clusters belonging to the CrB-SC. The 
X-ray luminosity and electron temperature are given for the four clusters included in 
the BCS catalogue \citep{ebeling_98a,ebeling_00a}; we have also included the more distant clusters 
A2069 and A2073, for which there are also data in BCS. Since we do 
not have information on the $\beta$ parameter (which describes the slope of the density profile in 
a $\beta$-model \citep{cavaliere_76}) and the core radius from X-rays, we have used the 
relation given by \citet{monteagudo_04} in terms of the X-ray luminosity ($L_x$), 
\begin{equation}
\Delta T_{SZ}^{RJ} = -(0.24\pm0.06) \left[\frac{L_{x}^{(0.1-2.4~keV)}}
{10^{44}~h_{50}^{-2}erg~s^{-1}} \right]^{0.47\pm0.09}~mK
\end{equation}
in order to make an estimate of the expected central SZ decrement at the VSA frequency
\[
(\Delta T_{SZ})_0 = -42.545\mu K \left[\frac{T_e}{keV} \right] 
   \left[\frac{n_{e0}}{10^{-3}cm^{-3}} \right] \left[\frac{r_c}{Mpc} \right]  
\]
\begin{equation}
 \times \int [1+ (r/r_c)^2 ]^{-3\beta/2}d(r/r_c) ~~,
\end{equation}
where $T_e$ is the electron temperature, $n_{e0}$ the central electron density and $r_c$ the 
core radius.

\begin{figure}
\includegraphics[width=\columnwidth]{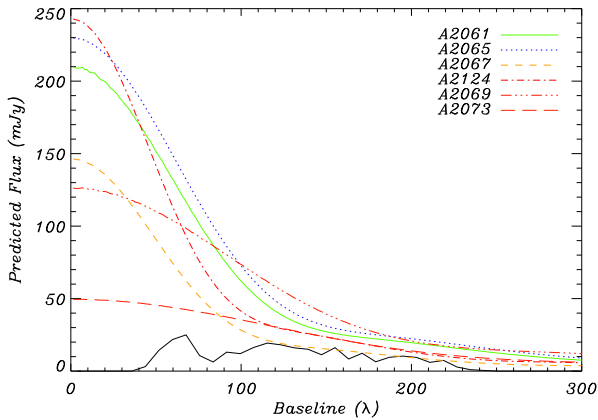}
\caption{Predicted SZ flux profiles in the aperture plane for four CrB-SC clusters plus other two 
clusters in the region,
as a function of the projected baseline length. The thick solid line represents the 
VSA weighting function (arbitrary units), computed from these observations.}
\label{profiles}
\end{figure}

Making use of the assumption that the inter-cluster (IC) medium is a perfect monoatomic gas in thermal
equilibrium, it follows that the temperature is proportional to $M^{\alpha}(1+z)$, with
$\alpha=\frac{2}{3}$, while the core radius is proportional to $M^{-\frac{1}{6}}(1+z)^{-\frac{1}{4}}$. 
Hence, we can obtain a scaling relation of the type 
$r_c=r_{c0}M_{0}^{-\frac{1}{6}}(T_e/T_{e0})^{-\frac{1}{4}}$, where $T_{e0}=10^8~K$, 
$M_0=10^{15}~M_{\odot}$, and $r_{c0}=0.13h^{-1}$~Mpc 
\citep{markevitch_98a}. These have been used to obtain a rough estimate of the core radius of
each of the clusters for which we have X-ray information. To convert to angular sizes, 
a cosmology $\Omega_M=0.25$, $\Omega_{\Lambda}=0.75$, $h=0.73$, as derived 
from the last VSA results \citep{rebolo_04}, was used. Making
use of these values, and assuming $\beta=\frac{2}{3}$, we can simulate the response of the 
VSA to each of these six clusters for which we have X-ray information. 
The brightness temperature map is converted into flux density,
multiplied by the VSA primary beam response, and Fourier transformed to obtain the simulated
aperture-plane response of the VSA to the SZ decrement. We must also consider the primary beam 
attenuation for clusters displaced from the pointing position. 
The cluster profiles in the aperture plane for the different 
observing baselines are shown in Figure~\ref{profiles}. It is worthwhile comparing this plot with
that presented in the VSA SZ work \citep{lancaster_04}, which shows
profiles with a higher amplitude, and similar shapes. Since, in that study, over a selected sample 
of rich Abell clusters and with similar noise levels, 
VSA data have shown clear SZ detections in five out of the seven observed clusters, we do not 
expect a high-significance detection of SZ effect from 
the single clusters in the CrB-SC region. Moreover, in this case the decrements are affected by the 
primary beam response, except in clusters A2019 and A2124, in which the pointing centres and cluster 
coordinates coincide. 

On the other hand, the simulated visibilities for each cluster, computed as
explained above, were convolved with the synthesised beam of the closest observation 
in order to measure the central flux decrements expected in the maps. 
These values are shown in Table~\ref{tab:clusters} under the symbol 
$\Delta S_{\nu}^{\rm pred}$.
If we compare the values with the uncertainty in our measurements, we expect to have 
SZ decrements from single clusters in the region typically with a confidence level between
$1$ and $2\sigma$, where $\sigma$ includes now the primordial CMB 
($\sigma_{CMB}\approx 19$~mJy/beam), the thermal noise ($\sigma_n$) and the residual 
sources ($\sigma_{sour}\approx 6$~mJy/beam) contributions. 
This is exactly what we obtain: comparing the cluster flux values from the final 
maps (see last column in Table~\ref{tab:clusters}), 
we find excellent agreement for those clusters that have a prediction for 
the flux. 
In all ten clusters analysed, 
we get negative flux values, and in the two most luminous clusters in the CrB-SC region, 
A2061 and A2065, we have $2\sigma$ detections. These two clusters are 
located close to larger decrements (see  Figure~\ref{fig:mosaic_zoom}). 
Moreover, according to Figure~\ref{fig:cor_plot} (which includes all galaxies lying within 
$\pm 0.02$ of the redshift reported in NED for each Abell cluster), A2065 is located 
in the region with the highest galaxy projected density. 
We can now combine all these individual measurements to obtain a 
statistical detection $-24.4 \pm 9.2$~mJy/beam ($2.7\sigma$).
If we also include in this weighted mean those values from A2069 and A2073,
we find  
$-24.8 \pm 8.1$~mJy/beam ($3.0\sigma$), for all the Abell clusters listed 
in Table~\ref{tab:clusters}.

Finally, another point worth noting is the presence of a $5\sigma_n$ (signal-to-noise level) decrement 
(decrement I, see Figure~\ref{fig:mosaic_zoom}) inside pointing CrB-I, located in a region with a 
large concentration of galaxy clusters, including A2089 and several Zwicky clusters. Their individual 
SZ effects could have an important contribution to its total decrement.
 
\begin{table*}
\begin{minipage}{150mm}
\begin{center}
\caption{Coordinates, flux densities and temperatures of the two largest decrements found in the mosaic. We 
quote both the thermal noise ($\sigma_n$) and, within parenthesis, the quadrature of the primordial CMB, 
thermal noise 
and residual sources ($(\sigma_{cmb}^2+\sigma_{n}^2+\sigma_{sour}^2)^{1/2}$) error bars. The last column 
indicate the percentage over the 5000 simulations with minima under each decrement flux density (within 
parenthesis, we also show this value when we apply a smoothing to the data; see Section~4.2.1 for details).}
\begin{tabular}{lccccc}
\hline
                 & RA (J2000)  &   DEC(J2000) & $\Delta S_{\nu}$ (mJy/beam) & $\Delta T_{RJ}$~($\mu$K)  & \%
		 below $\Delta S_{\nu}$  \\
\hline
Decrement B      & 15 25 21.60 & +29 32 40.7 &  $~-70\pm12~(\pm24)$   &  $-157\pm27~(\pm53)$ & 37.82 (54.16)\\
Decrement H      & 15 22 11.47 & +28 54 06.2 &  $-103\pm10~(\pm22)$   &  $-230\pm23~(\pm49)$ &  0.33  (0.74)\\

\hline
\end{tabular}
\label{tab:dec}
\end{center}
\end{minipage}
\end{table*}

\subsection{The origin of the negative spots}

The most remarkable feature of the mosaic is the existence of two prominent negative 
spots, at signal-to-noise levels of $6$ and $10$, situated inside the primary beam FWHM of 
pointings CrB-B and CrB-H, respectively. Both decrements are extended; indeed they 
cover an area equivalent to $\approx 3$ VSA synthesised beams. Coordinates, maximum negative 
flux densities and brightness temperatures of these strong decrements are listed in 
Table~\ref{tab:dec}. To convert the flux densities into brightness temperatures, we have 
used the Rayleigh--Jeans expression:
\begin{equation}
T_{RJ} = \frac{\lambda^{2}}{2 k_{B}\Omega_{sb}} S ~~,
\end{equation}
where $\lambda$ is the observing frequency, $k_B$ is the Boltzmann constant, and $\Omega_{sb}$ 
is the solid angle subtended by the synthesised beam.

These decrements are located in 
regions with no known clusters of galaxies (see Figure~\ref{fig:mosaic_zoom}). Decrement B is 
located near the centre of the supercluster, and slightly towards the north is the more 
distant Abell cluster A2069, which \citet{small_98a} have considered as a possible supercluster 
because of the large number of galaxies it contains. Decrement I and another negative feature 
(decrement J) located towards the east of the mosaic, inside pointing CrB-J, have 
similar flux densities to that of decrement B, but with a lower significance (signal-to-noise 
of $5$ in both cases). Note that this latter feature is located in the same position as the 
radio source 1532+2919, so the error bar of its measured flux ($51\pm12$~mJy, 
see Table~\ref{tab:sources}), may introduce an additional uncertainty.

We searched for possible negative features in the \textit{WMAP} maps \citep{bennett_03a} in the 
region of the most intense decrement. Given the angular resolution of the VSA extended 
configuration, we selected the \textit{WMAP} $W$-band (94~GHz, FWHM=$0.21\degr$) map.
From this, we can predict the observed visibilities as seen
by the VSA, enabling us to build a ``dirty'' map. Given the sensitivity of the \textit{WMAP} map
($\approx 170~\mu$K per $7$~arcmin pixel in this region, the corresponding 
error would be $\approx 108~\mu$K in a VSA synthesised beam of $11$~arcmin), 
we expect to see decrement H only marginally at the $2\sigma_n$ level. 
In the processed $W$-band map we do indeed see a negative spot in the region of decrement H in 
which the minimum temperature value, $\approx -146 \pm 108~\mu$K, is located at 
$15^{h}21^{m}19^{s} +28\degr 55' 27''$ (J2000). 
Note that although there is an offset of the 
observed centres of the spots  between the VSA and \textit{WMAP} of $\approx 10$~arcmin, 
this is consistent with the instrumental resolutions. Given the noise levels involved, we can
not conclude anything about the nature of the decrement, i.e. we cannot 
disentangle at $1\sigma_n$ level whether the spectrum between 
33~GHz and 94~GHz corresponds to primordial CMB 
(the decrement should have temperature $-230~\mu$K 
at 94~GHz in this case) or to SZ decrement (in which case the signal should have a 
temperature $-185~\mu$K at 94~GHz).

In principle, decrements B and H, given their sizes and amplitudes, could be either
extraordinarily large primordial CMB spots, or SZ signals related to either unknown clusters or 
to diffuse extended warm/hot gas in the supercluster. It should be noted that
only one third of the galaxies in the CrB region are linked to clusters, so 
SZ contributions could be expected in places where there are no catalogued 
clusters. As shown in Figures~\ref{fig:cor_plot}~and~\ref{fig:mosaic_zoom}, 
there are few known galaxies around decrement H, while decrement B is close
to a large concentration. This is important since
the position of galaxies could trace the warm/hot gas distribution (see, for example, 
\citet{monteagudo_04b}). 
In the next subsections we explore in detail the three possible explanations 
for the observed decrements.

\subsubsection{CMB anisotropy}

In order to quantify the possible contribution from the primordial CMB to these large 
spots, we carried out Monte-Carlo simulations. 
Using \textsc{cmbfast} \citep{seljak_96a} we generated a CMB power spectrum with a cosmological 
model defined by the following parameters:
$\Omega_B=0.044$, $\Omega_M=0.25$, $\Omega_{\Lambda}=0.75$, $h=0.73$, $\tau=0.14$, $n_S=0.97$, 
as derived from the most recent VSA results \citep{rebolo_04}, 
plus $T_{CMB}=2.725~K$, $\Omega_{\nu}=0$ \citep{bennett_03a}, and assuming adiabatic initial conditions. 
For each decrement we used this power spectrum to carry 
out 5000 simulations of VSA CMB observations, following the procedure explained in 
Section~3.3 of \citet{savage_04}, and using the aperture plane coverage from each pointings 
CrB-B and CrB-H as templates. Each visibility point contains the CMB and thermal 
noise contributions, plus the confusion level introduced by the residual radio sources below the 
subtraction threshold of $50$~mJy. Thus $V=V_{cmb}+V_{n}+V_{sour}$, where 
\begin{equation}
V_{sour} (\vec{u}) = \sum_{i=1}^{ns} {S_ie^{2\pi \vec{u}.\vec{x_i}}}~~,
\end{equation}
where $\vec{u}$ is the position vector of each visibility point, and $S_i$ and $\vec{x_i}$ 
the flux and the position vector in the map plane of the i-th source. 
The fluxes $S_i$ are generated using the source count derived from equation~\ref{sourcecounts}, 
yielding a total of $ns=894$ sources distributed between $1$ and $50$~mJy in a region within 
$2\degr$ of each pointing. The positions $\vec{x_i}$ are randomly and uniformly distributed 
inside this region.

In the last column of Table~\ref{tab:dec} we quote the percentage of realizations in which the 
minimum CMB flux value is below that found in the real map in the two cases in question. 
According to these results the probabilities of decrements B and H being caused by primordial 
CMB (adding the thermal noise and the residual sources components) are $37.82$\% and $0.33$\% 
respectively. Note that here we are considering only the intensity: the angular 
size of the spots is disregarded. In order to account for this, we have applied Gaussian 
smoothing (with a FWHM approximately equal to the FWHM of the spots) to both the simulated and the 
real maps, and in this case the probabilities of the decrements are higher 
(see also Table~\ref{tab:dec}). Hence, particularly for decrement H, the low probabilities are 
mainly due to the negative flux density values rather than the corresponding angular sizes.

We calculate the standard deviations of all pixels located within the $2.1\degr$-FWHM of the primary 
beam in the 5000 realizations, in order to estimate the confusion level introduced by the primordial 
CMB, the thermal noise and the residual sources components added in quadrature 
($\sigma=\sqrt{\sigma_{cmb}^2+\sigma_{n}^2+\sigma_{sour}^2}$, where we have 
$\sigma_{cmb}\approx 19$~mJy/beam, $\sigma_{n}\approx 11$~mJy/beam and $\sigma_{sour}\approx 6$~mJy/beam). 
From this, we find that decrements B and H are deviations at $2.95$ and $4.62\sigma$. These results 
clearly make a primordial CMB fluctuation alone an unlikely explanation for decrement H. We now 
focus solely on decrement H.

\begin{figure}
\includegraphics[width=\columnwidth]{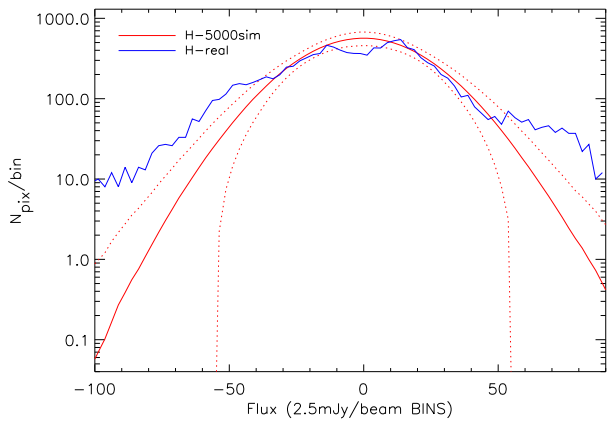}
\caption{Histograms comparing the distribution of flux densities inside the $2.1\degr$-FWHM of the 
primary beam in the 5000 simulations (thin curve) and in the real data of pointing CrB-H (thick curve). 
We also show the $1\sigma$ error bars of the simulations (dotted curve).}
\label{fig:hist_log_total_h}
\end{figure}

Following \citet{rubino_03} we performed a fluctuation analysis. In Figure~\ref{fig:hist_log_total_h} 
we present a logarithmic plot of the histograms (i.e. the $P(D)$ functions) of pixels values from the 
5000~realizations and also using the real data from pointing CrB-H. In the case of the realizations, the 
$1\sigma$ error bars (containing the primordial CMB, the thermal noise and the residual sources) 
are displayed. 
The plot shows a clear excess in the real data compared with the simulations at flux densities 
below $\sim -30$~mJy/beam, caused by the presence of decrement H. The excess in the positive tail 
of the curve comes from the two hot spots located towards the north and the east of 
decrement H in the final mosaic (see Fig.~\ref{fig:mosaic_zoom}). These positive features could be 
either real hot spots in the sky or artefacts due to the presence of decrement H itself.  
(Note also that the response of an interferometer has zero mean, 
so the presence of a strong negative feature in the map could enhance the neighbouring positive 
spots). It may be that these bright spots are enhanced 
by the presence of radio sources not identified in the extrapolation. However, these positive 
features are not present in the map of the region constructed using only VSA baselines 
$ > 125 \lambda$, suggesting that these structures are extended and not point sources.
In order to check if these positive spots are due to the sidelobes introduced by decrement H in the 
convolution with the synthesised beam, we \textsc{clean}ed the region and find that  
we obtain lower residuals by placing a \textsc{clean}-box around the 
negative spot than when the \textsc{clean}-box is placed around the positive spots. Furthermore,  
with the box around the negative hole, the intensity of the positive spot is reduced by $\approx 12$~\%. 
This is clear evidence that these bright spots are enhanced by the sidelobes 
of the synthesised beam. 
Using the \textsc{clean}ed map, the positive tail of the histogram is strongly reduced, while 
the negative tail is clearly dominant.

\begin{figure*}
\includegraphics[width=12cm]{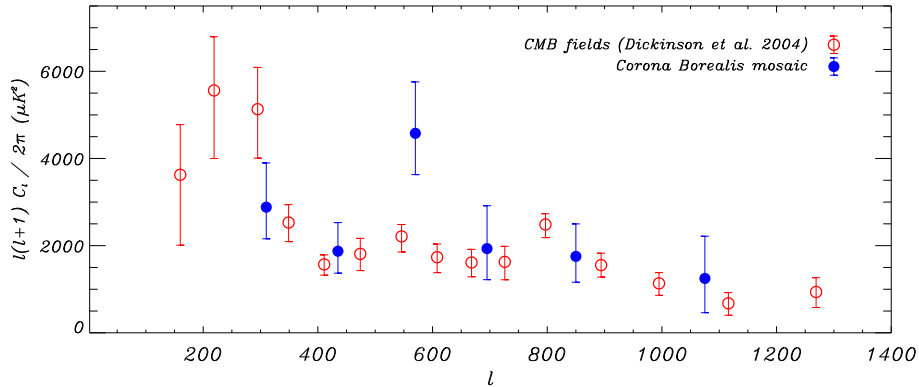}
\caption{Power spectrum computed from the CrB mosaic, in comparison with that obtained from the most 
recent VSA observations of primordial CMB fields \citep{dickinson_04}. The deviation at $\ell\approx 550$ 
is due to the presence of decrements B and H (see Section~4.2.1 for details).}
\label{fig:ps}
\end{figure*}

We also have computed the power spectrum of the mosaic, following the method described in 
\citet{scott_03}. This is shown in Figure~\ref{fig:ps}, along with that obtained from
the most recent primordial CMB observations with the VSA \citep{dickinson_04} for comparison. 
The comparison shows a $2.3\sigma$ deviation from the pure primordial CMB behaviour at 
$\ell\approx 550$. This is 
as expected, since this scale corresponds to the size of the large decrements. Indeed, we 
find that when we compute the power spectrum using data from pointing CrB-H alone, this offset increases 
to $3.0\sigma$. 
Moreover, as a consistency check, when we remove pointings CrB-B and CrB-H, which contain decrements B and 
H, the resulting power spectrum is then compatible with primordial CMB fluctuations on all scales. 

\subsubsection{SZ effect from clusters of galaxies}

In order to explore the possible SZ contribution, we
followed the formalism utilized by \citet{holder_00a} (see also \citet{kneissl_01} and 
\citet{battye_03}) to compute the number of clusters in this region potentially 
capable of producing a decrement like H. We estimate the mass a cluster would need to 
have in order to generate an SZ effect at least as intense as a given threshold. In this case 
the threshold is the amplitude of decrement H, although, to give an idea of the 
total number of clusters that would be detected in the region, we also considered the 
confusion level introduced by the quadrature sum of the primordial CMB, the thermal noise 
and the residual sources ($\sigma \approx 23$~mJy/beam). 
We then calculate the number of clusters per unit redshift in the Universe with 
masses above this in a solid angle equal to the entire surveyed region 
($\approx 24$~deg$^{2}$), by using the \citet{press_74a} (PS) mass function and also the 
\citet{sheth_99} (ST) mass function, as this provides a better fit to the 
simulations (note however that simulations also omit much physics). We have adopted 
$f_{gas}=0.12$ \citep{mohr_99a} for the gas mass fraction of 
the cluster, $\delta_c = 1.69$ \citep{peebles_80a} for the critical overdensity, 
$\sigma_8=0.84$ \citep{bennett_03a} for the spectrum variance in spheres of radius 
$8$h$^{-1}$~Mpc, and the same values as stated above for the other cosmological parameters.

\begin{figure}
\includegraphics[width=\columnwidth]{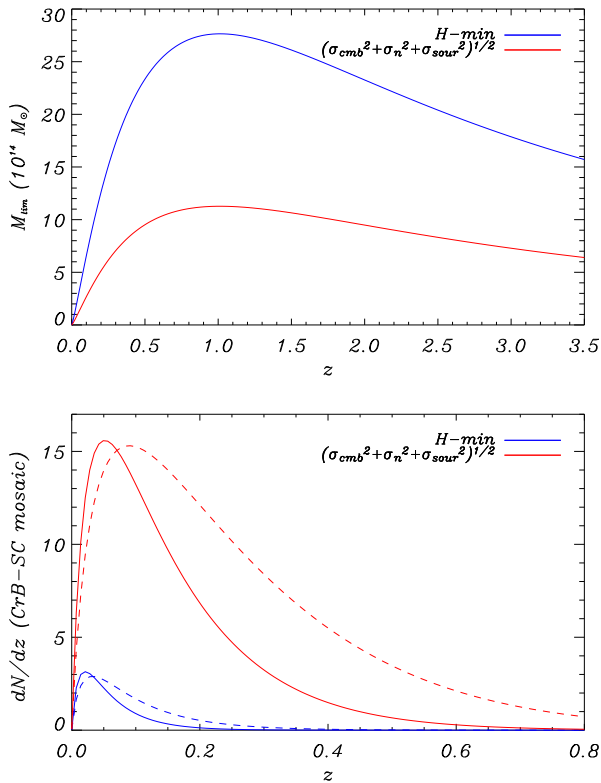}
\caption{Minimum SZ effect detectable mass versus redshift (top) and number of
expected clusters per redshift unit versus redshift (bottom), using as
thresholds either the sum in quadrature of the standard deviations of the primordial CMB, 
the thermal noise and the residual sources (thin curve), and the flux density of spot H 
(thick curve). The solid and dashed lines correspond respectively to the PS and ST mass functions.}
\label{fig:szprediction}
\end{figure}

\begin{table*}
\begin{minipage}{150mm}
\begin{center}
\caption{Maximum threshold masses, their redshifts, number of detectable clusters
in the whole survey, and the most likely redshift for clusters producing a SZ effect 
larger than either the confusion level introduced by the primordial CMB, the thermal 
noise and the residual sources and the minimum flux density of pointing H.
We have used both the PS and the ST mass functions.}
\begin{tabular}{|c|c|c|c|c|c|c|}
\hline
Threshold & Maximum  & z where $M_{lim}$ & \multicolumn{2}{c}{Total number} & \multicolumn{2}{c}{z
where $dN/dz$} \\ 
 &$M_{lim}$ ($10^{14}M_{\odot}$)&is maximum  & \multicolumn{2}{c}{of clusters} &
 \multicolumn{2}{c}{is maximum} \\
      &    &    &   PS   &   ST  &   PS    &   ST   \\
\hline
 $|$Min-H$|$=$103$~mJy/beam   & $27.65$ & $1.01$ & $ 0.3$ & $ 0.4$ & $0.021$ & $0.035$ \\
 $\sqrt{\sigma_{cmb}^2+\sigma_{n}^2+\sigma_{sour}^2}$ $\approx$ $23$~mJy/beam & $11.27$ & $1.01$ & $ 3.2$ & $ 5.4$ & $0.049$ & $0.091$ \\
\hline
\end{tabular}
\label{tab:szprediction}
\end{center}
\end{minipage}
\end{table*} 

Figure~\ref{fig:szprediction} shows the derived threshold mass and the number of clusters per 
unit redshift versus redshift. By integrating this last curve we obtain the total number of 
expected clusters in the surveyed region, which is presented in Table~\ref{tab:szprediction}.
The number of SZ clusters that should be detected above the confusion level introduced by the 
primordial CMB, the thermal noise and the residual sources in the whole surveyed region is 
$3$ and $5$ respectively for the PS and the ST mass functions. They should be located at mean 
redshifts $0.049$ or $0.091$, which is close to the mean redshift of the CrB-SC 
($\approx 0.07$). 
Note that in our maps we have detected three clusters in the region above this confusion level 
(see Table~\ref{tab:clusters}). 
On the other hand, the number of clusters that could produce a decrement at least as intense as
H is respectively only $0.3$ or $0.4$ for the PS and the ST prescriptions. Therefore, the 
probability of decrement H of being caused entirely by a single cluster is very low. 
However, if we take into account that a decrement such as H due to SZ may be enhanced by the 
primordial CMB, this probability must be higher.
Note finally that these statistics are valid only for a \textit{random} patch of sky. 
As we are observing a selected overdense region in the direction of a supercluster, the actual 
probabilities \textit{must} be higher. 
To obtain a more precise result we may turn to N-body simulations, although our analysis 
is sufficient to get an order of magnitude estimation, and to show that these probabilities are low.

\subsubsection{Diffuse extended warm/hot gas} 

As indicated in Section~1, superclusters may be reservoirs of
diffuse warm/hot gas, which in principle might produce a detectable imprint 
in the low-energy X-ray bands. \citet{zappacosta_02a}, and more recently 
\citet{zappacosta_04a}, have claimed an excess of X-ray emission 
from diffuse structures in a high galactic latitude \textit{ROSAT} field, and the detection 
of a significant correlation between the \textit{ROSAT}-PSPC pointings and the M\"{u}nster 
Redshift Survey of galaxies in the Sculptor supercluster region. 
The SZ effect may provide another tool to detect this gas, thanks to the long 
path lengths of the photons across these large-scale structures. 
Using the \textit{COBE}-DMR data, \citet{molnar_98} were unable to find evidence 
of an SZ imprint in the region of the Shapley supercluster.  

\begin{figure}
\includegraphics[width=\columnwidth]{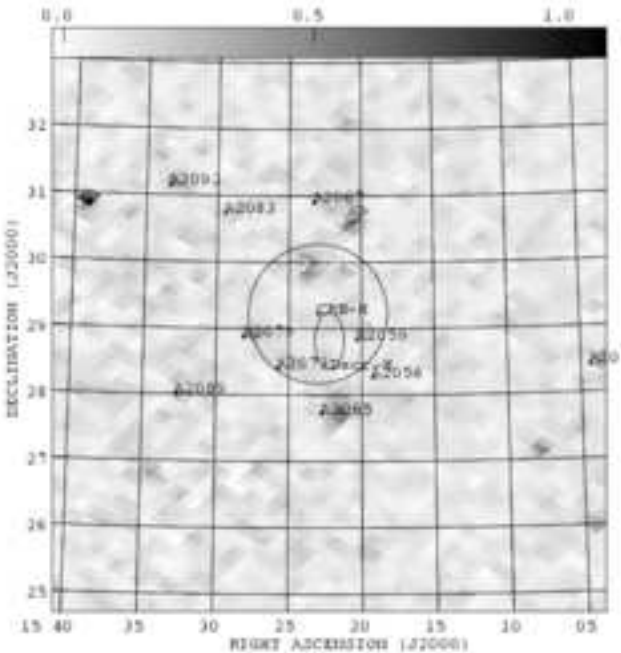}
\caption{\textit{ROSAT}--R6 (0.73-1.56~$keV$) \citep{snowden_97} map centred on pointing CrB-H. 
We indicate the positions of all the Abell clusters in the region, emphasizing the CrB-SC clusters 
with a star. The units indicated in the upper flux bar are $10^{-3}$~counts/s/arcmin$^2$. The 
circle indicate the FWHM of the primary beam of pointing CrB-H, while the ellipse 
corresponds to the position of decrement H in the VSA mosaic. 
There is no clear excess of emission in this region.}
\label{fig:R6}
\end{figure}

Since decrement B, according to our studies, can be explained through a  
primordial CMB decrement, we shall focus on decrement H, and explore whether it could  
have been built up from a concentration of diffuse warm/hot gas ($T_e<1~keV$).  
Such a structure producing a detectable SZ effect could also leave an imprint 
on the less energetic X-ray bands.
We have analysed the \textit{ROSAT} XRT/PSPC All-Sky Survey \citep{snowden_97} 
map, corresponding to the R6 band (0.73--1.56~keV), in order to search for 
correlated X-ray diffuse emission in the region.
The R6 map does not show an excess of emission in the
position of the decrement (see Figure~\ref{fig:R6}). The signal in the 
region of the decrement is $\approx 80-140\times10^{-6}$~counts/s/arcmin$^2$, 
which comes mainly from the background.  
In order to perform the \textit{ROSAT}-VSA correlation,  we have adopted two
different methods.

\begin{figure*}
\includegraphics[width=14cm]{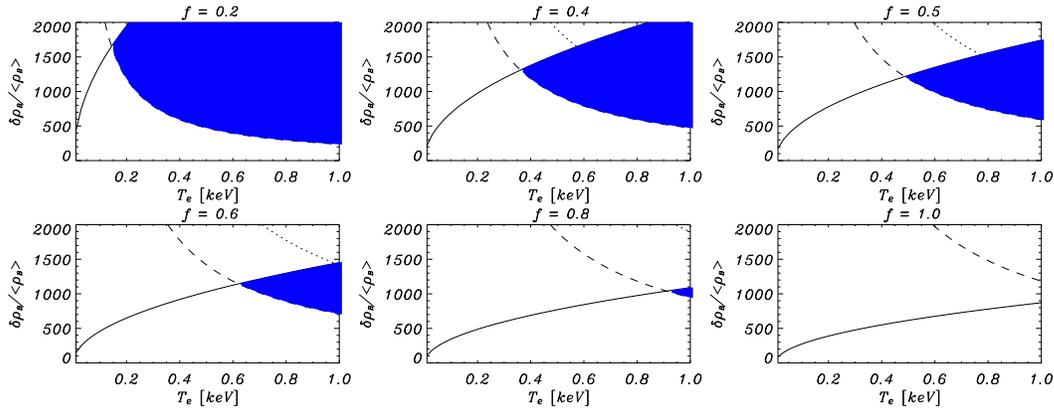}
\caption{Constraints placed on the overdensity (expressed in terms of baryon density over 
the mean baryon density) and the electron temperature 
in the centre of the assumed gas concentration (see Section~4.2.3 for details) that generates 
decrement H. We take into account that the X-ray emission must be below that in 
the \textit{ROSAT}--R6 map (solid line), and that the depth of this structure should be lower 
than $40$~Mpc (dashed line), or lower than $20$~Mpc (dotted line). Filled regions indicate 
the acceptable combinations in the parameter space (for $40$~Mpc). The factor f represents the 
fraction of the decrement that is produced by the SZ effect.}
\label{fig:accept_parameters}
\end{figure*}

Firstly, we have applied a pixel-to-pixel comparison of the
\textit{ROSAT}-R6 map and the real space \textsc{clean}ed mosaic
following the method described in \citet{monteagudo_04}.
This method considers the brightness temperature measured by a 
CMB experiment such as the VSA as the sum of different components: 
a cosmological signal $T_{CMB}$, the template we want to measure
$M$ (in this case, the thermal SZ component traced by \textit{ROSAT}),  
instrumental noise $N$, and foreground residuals $F$.
The total signal measured at a given position on the sky
is then modelled as $T=T_{CMB}+\alpha M+N+F$, where $\alpha$ measures the
amplitude of the template induced signal.
If all the other components have zero
mean and well known correlations functions, and ${\cal C}$ denotes the
correlation matrix of the CMB and noise components, then the estimate of
$\alpha$  and its statistical error are
\begin{equation} 
\alpha = \frac{ T {\cal C}^{-1} M^T} {M {\cal C}^{-1} M^T }, 
\;\;\;\; \sigma_\alpha = \sqrt{\frac{1}{M{\cal C}^{-1}M^T}}  ~~.  
\label{eq:alpha1} 
\end{equation}
The derived value of the correlation between \textit{ROSAT} and the VSA mosaic is
$\alpha=-0.28\pm0.74~\mu K/X$ ($X$ denotes the units of the 
R6 map, which are counts~s$^{-1}$~arcmin$^{-2}$), 
which means that there is no significant
anticorrelation, as would be expected if it were the case that the decrement is 
produced by a diffuse gas distribution. 

We repeated this analysis but in visibility space,
by predicting the expected visibilities from the \textit{ROSAT} map as seen
by the VSA, and performing a similar comparison between the observed and
predicted values. 
In this case, we find $\alpha = -0.21 \pm 0.57~\mu K/X$ for the 
CrB-H field. The significance of this value was derived by computing the 
dispersion of the $\alpha$ value when performing a series of rotations of
one map with respect to the other prior to the computation of the predicted
visibilities.

Assuming the distance of the supercluster ($\approx 200h^{-1}$~Mpc, derived from 
the average redshift of the member clusters $\langle z\rangle=0.0725$),  
the physical transverse size of decrement H is $\approx2h^{-1}$~Mpc. At 
typical WHIM electron temperatures , i.e. $T_e = 0.01~keV$ to $T_e = 1~keV$, a 
homogeneous and spherical gas 
distribution of this size producing a central SZ decrement of $230~\mu$K would 
have such a high $n_e$ that it must have collapsed, with subsequent 
virialisation. This should also produce detectable X-ray emission, which is 
not present in the \textit{ROSAT}-R6 map. 
In a similar case of a CMB decrement without 
X-ray emission, \citet{kneissl_98} discuss various 
possibilities, but find it difficult to reconcile observations 
with the process of structure formation. A filament pointing towards 
us seems in some ways one of the more attractive options. 
We may therefore consider a larger structure with 
a lower density so that the path length is long enough to produce a detectable SZ 
effect ($\Delta T_{SZ}\propto\int n_{e}T_{e}dl$) without significant X-ray 
emission ($S_{X-ray} \propto \int n_{e}^2 T_e^{1/2} dl$). 
We can estimate the peak SZ intensity caused by 
this hypothetical filamentary structure aligned along the line of sight, with a depth $L$, 
a central electron density $n_e$, and electron temperature $T_e$, at the VSA frequency of 
$33$~GHz:
\begin{equation}
\Delta T_{0} f = -21.3~\mu K~ \left[ \frac{L}{Mpc} \right ]  
\left[ \frac{n_e}{10^{-3}cm^{-3}} \right ] \left[ \frac{T_e}{keV} \right ]  ~~.
\label{SZ0}
\end{equation}
Here, $\Delta T_{0}=230~\mu$K is the peak SZ decrement. Given that there could also be 
a contribution from primordial CMB anisotropies, we have introduced the factor $f$ which 
accounts for the fraction 
of the decrement due to the SZ effect. For simplicity, we have assumed isothermality 
and that the electron density is homogeneous along the line of sight. We have 
assumed that the depth of this structure must be lower than the maximum separation 
along the line of sight between clusters in the core of the CrB-SC, which is 
$29h^{-1}$~Mpc = $40$~Mpc.  
These assumptions set constraints on the overdensity (expressed by the baryon 
density over the universal mean baryon density)  
and $T_e$ which are represented in Figure~\ref{fig:accept_parameters} by 
the dashed ($L<40$~Mpc) and the dotted lines ($L<20$~Mpc).

We also require the Bremsstrahlung emission to be low 
enough that the signal does not leave a detectable imprint on the \textit{ROSAT}-R6 map. We  
use equation~\ref{SZ0} to derive a relation between $L$ and $T_e$, and integrate along the 
line of sight. We consider a background level of $10^{-4}$~counts/s/arcmin$^2$ 
(see Figure~\ref{fig:R6}) as 
an upper limit for the X-ray emission in the region of decrement H. Scaling with the signal 
of the Coma cluster in the same band ($63\times10^{-4}$~counts/s/arcmin$^2$), 
we obtain the constraint represented by the solid line of Figure~\ref{fig:accept_parameters}. 
Filled zones in the parameter space show the acceptability regions for 
structures with lengths of less than $40$~Mpc. These plots show that for $f=1$ there is no   
reasonable combination of $n_e$ and $T_e$ to explain the whole decrement. Only 
values of $f\la 0.7$ are able to explain the decrement. 
For instance, a filament with a temperature $T_e \approx 0.7$~keV, a depth 
$L\approx 40$~Mpc, and a baryon overdensity 
$\delta\rho_B/\langle\rho_B\rangle \approx 850$ would produce 
a X-ray signal of $\approx 0.6\times10^{-4}$~counts/s/arcmin$^2$, which is almost a 
factor $2$ below the background confusion level of the \textit{ROSAT}-R6 map. 
A structure having these parameters would also produce a peak SZ effect of 
$\approx 115~\mu$K, which is half of the amplitude of the decrement ($f=0.5$). Also, a 
remarkable \textit{alignment} is required to achieve this fraction of brightness contrast.  

If instead of the value 
$\delta\rho_B/\langle\rho_B\rangle\approx 850$, which is computed for the centre of the 
filament, we assume an average value of $\delta\rho_B/\langle\rho_B\rangle \approx 450$ over 
the entire structure, and a $2\times 2$~Mpc$^2$ square for the transverse shape of the filament, 
the total gas mass enclosed in it would be $M_{gas}\approx 5\times10^{14} M_{\odot}$. 
This overdensity appears to be difficult to explain in the light 
of current hydrodynamical galaxy formation simulations, which predict 
overdensities $\approx 10$ times lower for these large-scale structures \citep{cen_99a}. 
If we assume that the dark matter is well traced by the baryonic matter, this leads to an 
overdensity of $\approx 110$ in terms of the total matter, in contrast to the typical values for 
non-bound supercluster scales with $\delta\rho/\rho_c \sim 5-40$, for structures that have just 
become virialised with $\delta\rho/\rho_c \sim 200$, or inside the typical Abell clusters with 
$\delta\rho/\rho_c \sim 1000$. We note that the derived gas mass contained in this hypothetical 
filament is comparable to the total baryonic mass in the known CrB-SC clusters, and would 
represent $\approx 10$\% of the total baryonic mass of the supercluster. 

\section{CONCLUSIONS}

We have \textit{selected} the wholly unusual Corona Borealis supercluster region for CMB observations 
with the VSA extended array at $33$~GHz. The structures detected in the mosaiced map of the core of 
the supercluster have significant contributions from primordial CMB anisotropies. 
However, these maps show negative flux values at the positions of the ten richest clusters in the 
region, and the two most X-ray luminous CrB-SC clusters, A2061 and A2065, each produce $2\sigma$ 
decrements. When we combine the flux values from the positions of the CrB-SC member clusters, we 
obtain a statistical SZ detection with a significance of $2.7\sigma$ ($24.4\pm9.2$~mJy/beam). 
If the clusters A2069 and A2073, which have higher redshifts, are added to the sample, we obtain 
$3.0\sigma$ ($24.8\pm8.1$~mJy/beam).  
In addition to this, the mosaic shows, in a region with a high density of Zwicky clusters, 
a large negative feature, at a signal-to-noise ratio of 5, which could be enhanced by the SZ effects 
generated by these single clusters.

This mosaic shows two strong and extended negative features with flux densities $-70\pm12$~mJy/beam 
($-157\pm27 ~\mu$K), and $-103\pm10$~mJy/beam ($-230\pm23~ \mu$K), and positions 
$15^{h}25^{m}21.60^{s} +29\degr 32' 40.7''$ (J2000) and 
$15^{h}22^{m}11.47^{s} +28\degr 54' 06.2''$~(J2000), respectively. 
These decrements are located at positions where there are no known clusters. The first one  
is placed near the reported centre of the supercluster. In order to disentangle their 
origins, we have considered the possibility of large primordial CMB fluctuations, or 
SZ signals related to either unknown clusters or to diffuse extended warm/hot gas in the intergalactic
medium of the supercluster. The primordial contribution has been explored by performing simulations 
of primordial CMB observations with the VSA. 
These have shown that the size and intensity of decrement B are consistent with primordial CMB 
fluctuations, whereas in the case of decrement H the probability is only $0.33$\%. 
Decrement H is marginally detected in the W-band map of the \textit{WMAP}'s first-year dataset, but the current 
sensitivity does not allow us to disentangle the spectral behaviour of this decrement.

In order to investigate the second possibility we have predicted the number of clusters capable of 
producing a decrement such as H. To this end we used the Press-Schecter (PS) and also the 
improved Sheth-Tormen (ST) prescriptions to describe the density of collapsed objects in the Universe. 
This study showed that the number of random clusters in the VSA observed fields massive enough 
to produce this feature is 0.3 and 0.4 respectively for the PS and the ST formalisms. 
Note that this probability has been computed assuming that the whole decrement 
is due to the SZ effect. If we take into account that there could be a primordial CMB contribution, 
this probability becomes higher. 

The \textit{ROSAT}-R6 data show no evidence of X-ray emission in the region of decrement H. This fact sets 
constraints on the electron density and temperature that could have a hypothetical intercluster warm/hot 
gas distribution capable of producing a decrement as deep as H. If it were 
in a small volume, then the electron density would be high enough to produce 
detectable X-ray emission in the typical ranges of WHIM temperatures ($0.01-1$~keV). 
Hence, we hypothesise that the decrement could be caused by a large filament 
aligned in the direction of our line of sight. 
We have explored the possibility that only some fraction of the spot is 
due to an SZ effect, the primordial CMB being responsible for the rest. In this case, a filament with a 
length below the size of the supercluster, i.e. $\la 40$~Mpc, a temperature of $0.5-0.8$~keV, and a 
baryon density in its centre between $500$ and $900$ times the mean baryon density in the local Universe, 
produces a SZ effect close to one half 
the central flux density of the decrement, and its X-ray emission would be 
low and obscured by the background in the \textit{ROSAT}-R6 map. It would contain a gas mass of 
$\sim 5\times 10^{14} M_{\odot}$, which is comparable to the 
total baryonic mass contained in the CrB-SC member clusters. 
If the value $3\times10^{16} M_{\odot}$ is assumed for the total mass of the supercluster, this filament 
would hold the $\sim 10$\% of the total expected baryonic mass of the supercluster. 
However, the required overdensity of such structure is $\sim 10$ times higher than the predictions 
from N-body galaxy formation simulations for these large-scale filaments. But we stress that: i) we are 
observing at a very atypical region, and ii) N-body simulations need more physics.

In summary, we are confident that our measurements do show excess decrement, and 
to explain decrement H we require a combination of primordial CMB fluctuations with 
either an SZ effect from an unknown cluster or from a large-scale filamentary structure, which would hold a 
significant fraction of the total baryonic mass of the supercluster.
It is worthwhile to carry out multi-frequency observations of decrement H in order to disentangle 
these possibilities. 

\section*{Acknowledgments}
We thank the staff of the Teide Observatory, Mullard Radio Astronomy Observatory and Jodrell Bank
Observatory for assistance in the day-to-day operation of the VSA. We thank PPARC for funding and
supporting the VSA project. Partial financial support was provided by Spanish Ministry of Science and
Technology project AYA2001-1657. We acknowledge E. Battistelli, F. Atrio-Barandela and 
J. Betancort-Rijo for useful comments and discussions. CD is funded by a Stanley Rawn post-doctoral 
scholarship. 
We acknowledge the use of the NASA/IPAC Extragalactic Database (NED), operated
by JPL (Caltech), under contract with NASA. We also acknowledge the use of the \textsc{aips} package, 
developed by the NRAO.

\bibliography{coronab_paper}
\bibliographystyle{mn2e}


\bsp 

\label{lastpage}

\end{document}